# Design for Distributed Moroccan Hospital Pharmacy Information Environment with Service Oriented Architecture

Hajar Omrana, Safae Nassiri, Fatima-Zahra Belouadha and Ounsa Roudiés
Siweb Team (Information System and Web), SIR Laboratory
Ecole Mohammadia d'Ingénieurs, Mohammad V University- Agdal
Rabat, Morocco
homrana@emi.ac.ma, safaenassiri@gmail.com, belouadha@emi.ac.ma, roudies@emi.ac.ma

*Abstract*—In the last five years, Moroccan e-health system has focused on improving the quality of patient care services by making use of advanced Information and Communications Technologies (ICT) solutions. In actual fact, achieving runtime and efficient information sharing, through large-scale distributed environments such as e-health system, is not a trivial task. It seems to present many issues due to the heterogeneity and complex nature of data resources. This concerns, in particular, Moroccan Hospital Pharmacy Information System (HPIS) which needs to interact with several disparate medical information systems. Service Oriented Architecture (SOA) offers solution that is both flexible and practical to effectively address the problem of interoperability of e-health systems. In this paper, we discuss the limits and challenges of the current Moroccan information system intended for hospital pharmacy. We therefore propose a global Web services-based e-health architecture for integrating different heterogeneous blocks and various data resources of this system. We also present a solution to secure Web services communication using *WS-SecurityPolicy*.

*Keywords- Moroccan e-health; Hospital Pharmacy Information System; Security; WS-SecurityPolicy; Business Intelligence*

## I. INTRODUCTION

Nowadays, promoting health becomes a national priority of many countries through the world. In this context, e-Health area aims to benefit from advanced Information Technology (IT) to produce efficient healthcare systems able to communicate in a distributed environment. To achieve this goal, it is necessary to use appropriate mechanisms and technologies in order to enable managing and sharing huge medical information between all healthcare organizations and authorities, as well as providing efficient public health services. Within this framework, Moroccan government is reconsidering its current healthcare systems in order to overcome their limitations.

Recently, Moroccan Hospital Pharmacy Information System (HPIS) becomes a key concern for government e-health strategy. This system is used to guarantee supply and management of drugs over all Moroccan regions. However, it has some limitations. Heterogeneous regional systems and data resources impose integration issues that make it difficult to hold down drugs supply costs. To address this problem, we think that an efficient solution should be built on the use of Service Oriented Architecture. This architecture, known as a prominent solution for systems interoperability, will allow a smooth collaboration between all partners involved in the drug supply's process, independently of used platforms. The use of XML-based standards as SOAP (Simple Object Access Protocol) [1] and WSDL (Web Services Description Language) [2] enable all participants to easy communicate and share data. This might, in general, significantly reduce costs and time consuming when performing the global drugs management process.

Thereby, the main objective of this work is to propose a global Web services-based e-health architecture which essentially focuses on Moroccan HPIS' Web services, in order to ensure interoperability and data exchange between healthcare authorities and heterogeneous hospital pharmacies' systems. This paper is an extended version of the work [3] presented in ICID 2011 conference. The additional content proposed in the present work comparing to the ICID2011 paper concerns the use of *WS-SecurityPolicy* [4] to enable publishing security policies.

Remainder of this paper is structured in four sections. Section 2 presents Moroccan HPIS and exposes its limitations. Section 3 describes the proposed architecture and section 4 concludes this work.

## II. E-HEALTH AND SERVICE ORIENTED ARCHITECTURE

Public e-Healthcare has gained significant interest from the research community in recent years. In literature, several works [5-8] propose the use of Service Oriented Architecture (SOA) to bring more efficient solutions to e-Health open issues. This architecture enables the construction of collaborative services with higher reusability, flexibility, extensibility, and robustness [9] which can exchanges data and information and process tasks on the network.

The current architecture of medical systems in Morocco presents some limits in communicating and exchanging data between different bricks of the overall system. This impacts the quality of e-health system, reduces its performance and make difficult to meet the patient needs. To overcome these issues, we propose the use of *de facto* implementation of SOA paradigm: Web services, to package medical functionalities into services and allow performing efficient processes and data integration within Moroccan medical systems.

The key benefits of using SOA architecture in e-health context are the following:
- Creating collaborative environment independent of software platforms
- Enabling the reuse of existing healthcare systems
- Offering flexibility in interaction medical process and exchanging data

### III. MORROCAN HPIS: CURRENT LIMITATIONS

Moroccan HPIS is based on a drugs supply process that involves four main actors :

- **Department of Drugs Supply:** It is a ministerial department which annually conducts an assessment of pharmaceutical's national needs, ensures the drugs purchasing from drugs suppliers and delivers drugs to all regional healthcare structures.

- **Regional Depot of Drugs:** It receives manually signed drugs' orders from hospital pharmacies, delivers the ordered quantity to the hospitals, and manages the regional drugs' stock. In case of stock's rupture, the department sends a manually signed drugs order to Department of Drugs Supply.

- **Hospital Pharmacy System:** Each hospital has its own pharmacy's system which manages the hospital drugs' stock, sends manually signed drugs orders to Regional Depot of Drugs and stores prescription data in its database.

- **Drugs Supplier:** This is either a national or international supplier which receives manually signed drugs orders from Department of Drugs Supply, and delivers the ordered quantity to it.

Figure 1 summarizes the interactions among all present drugs supply process's actors described above.

Besides, the actual Moroccan HPIS presents several limitations. First, the systems and platforms used by the involved actors are completely heterogeneous and geographically dispersed. As a result, data sharing and communication between these systems are actually very difficult, costly and time consuming. A lot of time and effort are lost in sending manually signed drugs orders. Therefore, Regional Depot of Drugs receives drugs orders from a hospital pharmacy in extended delay. Moreover, Department of Financial Healthcare faces the same issues when receiving drugs orders from regional departments or ordering drugs from suppliers. In addition, the managed healthcare data is complex. In medical context, pharmacists, doctors and administrators may use different technical terms referring to the same health concept (e.g. heart attack and myocardial infarction). To address all of these issues, those systems may need to collaborate together using e-business process integration solutions. This must be done in accordance with security standards, Web semantic techniques interoperability and scalability mechanisms to ensure efficient applications interoperability and data exchange.

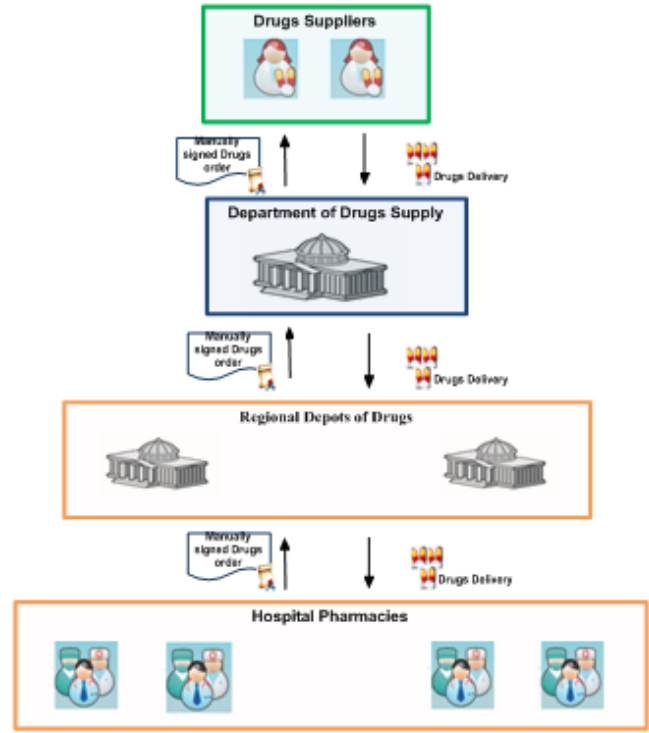

Figure 1. Interactions within Drugs Supply Process

Besides, as mentioned above, Department of Drugs Supply annually performs an assessment of pharmaceutical national needs by collecting prescriptions' data from disparate resources. The aim is to define the precise drugs quantity which must be delivered to each region. However, the decision concerning this issue is not easy to make. In fact, it is not actually possible to easy calculate indicators giving information about drugs' regional consumption. Each hospital pharmacies store patient's prescriptions in their own local databases using heterogeneous data structures. Therefore, extracting related data remains a hard task. To optimize the drugs distribution process, we think that it will be necessary to use an appropriate system architecture making easy the extraction of the required indicators.

### IV. PROPOSED MOROCCAN HPIS ARCHITECTURE

To achieve the goal of the global national e-health strategy that intends to provide an interoperable, standardized and secure platform for all involved partners in supporting healthcare services, our approach aims to propose a global Web services-based distributed architecture for Moroccan e-health. In this paper, we are in particular interested in the case of HPIS. Our contribution should remedy to all of the limitations noted above. Figure 1 illustrates the proposed architecture and shows how it fits into our global national e-health infrastructure vision. This architecture takes advantage of Web services technologies in order to allow performing efficient processes and data integration within Moroccan HPIS. This enables effective collaboration among current and future healthcare services the global e-health system. Besides data interoperability, our proposal also considers crucial aspects, such as security, semantic interoperability and scalability.

In the following sub-sections, we give an overview of the design and essential technical details related to the proposed architecture. We first present the key capabilities of e-health platform and then expose the proposed Web services that enable drugs management process integration and prescription data extracting.

*A. e-Health architecture key capabilities*

We think that there are two essential key capabilities to consider when designing e-health architecture: systems and semantic interoperability. To solve the problem of e-health systems interoperability and data exchanging, the designed architecture should consider a standardized way to make applications collaborate through Internet. The traditional business process integration solutions, conceived in this context, are costly systems which also use specific protocols and data formats [10]. That's why we propose the use of Web services which offer a flexible solution ensuring effective collaboration between software systems [11]. This technology indeed presents several advantages. First, it is independent of any particular platform or programming language. Second, Web services standards (e.g., SOAP, WSDL and UDDI) are XML-based technologies. Last and not least, Web services-based integration is low-cost, and constitutes a rapid development solution.

Besides, in distributed systems such as e-health system, there is a need to exchange human and machine understandable information in order to achieve full interoperability. To guarantee semantic interoperability, we propose the design of a shared national e-Health ontology based on OWL 2 [12] standard. This ontology must include concepts and properties related to the health domain, as well as relationships between those concepts. It would be useful to semantically annotate e-health Web services using SAWSDL standard [13] that is an extension of WSDL supporting Web services semantic aspects

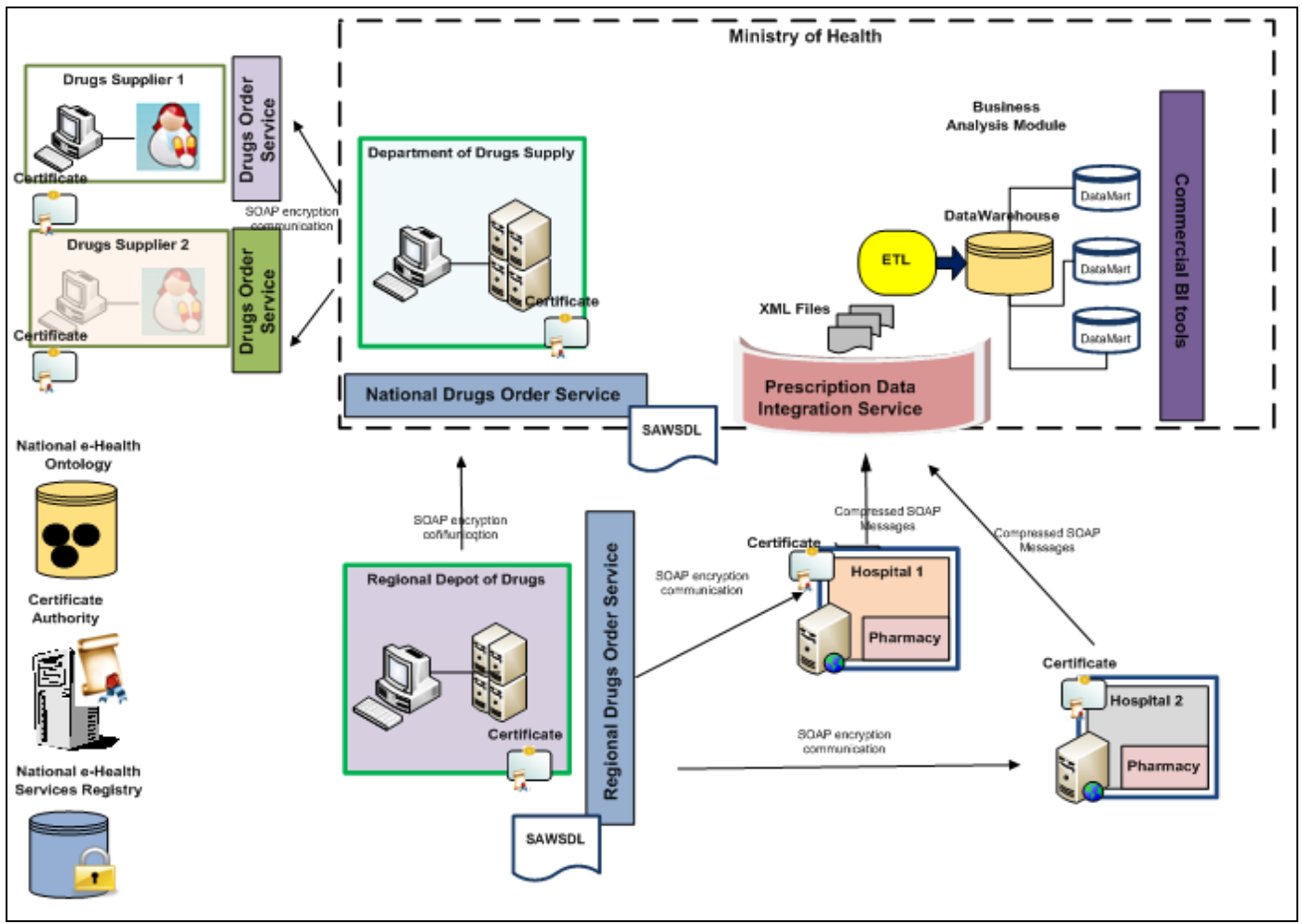

Figure 2. Web Services-based Moroccan Hospital Pharmacy Information System

### B. Moroccan HPIS Web services

To enable Moroccan HPIS interoperability, we propose to develop National and Regional Drugs Order Services. Each regional depot's system should develop *Regional Drugs Order Service* in order to instantly receive the drugs orders of all hospital pharmacies' systems belonging to its region. As for Department of Drugs Supply's system, it should implement *National Drugs Order Service* in order to interact with depots' systems. Besides, we also propose to implement Prescription Data Integration Service in order to optimize drugs distribution.

The two National and Regional Drugs Order Services are developed to enable medicines supply business processes' integration in a collaborative environment, respecting the global e-health infrastructure. They are referenced in *National e-Health Services Registry* and semantically annotated using SAWSDL standard and *National e-Health Ontology*. To be able to interact with these services, each hospital pharmacy system would have a Regional Drugs Order Web service Client stub. When the hospital pharmacy administrator stores a drugs' order record in the local system, Regional Depot of Drugs can be instantly alerted through secured SOAP messages. Similarly, Department of Drugs Supply will be alerted if national drugs' orders are submitted.

Concerning the drugs distribution issue, we recall that hospital pharmacy systems, private pharmacies or clinics, store patients' prescriptions in their own databases using heterogeneous technologies and platforms. Extracting prescription's data from these various and scattered data Security solution sources in order to be used for analyzing the national drugs' consumption, is actually a complex task. In our proposal, we use Web services technology to dynamically integrate data from involved data sources into XML files. That's why we use *Prescription Data Integration Service*. When the required XML files are available, An ETL component will be then able to extract necessary data from those files, to transform it according to the defined rules, and to load it into a system data Warehouse for business analysis considerations. To guarantee efficient data transmission, we adopt SOAP compression mechanisms. We use a Servlet filter to compress both request and response SOAP messages. The filter has to be added to the Axis Web application containing the developed Web Service. Finally, *Prescription Data Integration Service* is especially designed to resolve current prescription' data integration issues.

### C. Secured HPIS Web services using WS-SecurityPolicy

Given the sensitivity of information exchanged, e-health systems require a high level of security. For example, the patient's record information should be viewed by only the treating doctor or the patient himself. Similarly, this information should be treated or updated, by only the authorized profiles.

The proposed e-health architecture is a Web-based architecture. Therefore, it inherits risks and vulnerabilities existing in the Web. Security measures should be implemented to secure SOAP messages and XML exchanges.

HPIS Web services might have different security policies. Each Web service's provider could specify its own policy. To provide enough information for Web service consumers to be engaged in a secure exchange of messages, we propose, the use of *WS-SecurityPolicy* [4].

*WS-SecurityPolicy* is a Web services specification that offers mechanisms to represent the security capabilities and requirements of web services as policies. Security policy assertions are based on the *WS-Policy* framework [14]. *WS-Policy* is a W3C recommendation published in 2007. It constitutes a general formalism, based on XML, used to describe the policies, adopted by Web Services providers, and required by the client as well. *WS-SecurityPolicy* defines a set of security policy assertions for use with the *WS-Policy* framework with respect to security features provided in *WSS: SOAP Message Security*, *WS-Trust* and *WS-SecureConversation*. These assertions describe how messages are to be secured, using cryptographic algorithms and digital signature mechanisms.

SAWSDL files used to describe the HPIS Web services integrate *WS-SecurityPolicy* tags that inform about the security policy of the providers (figure 3).

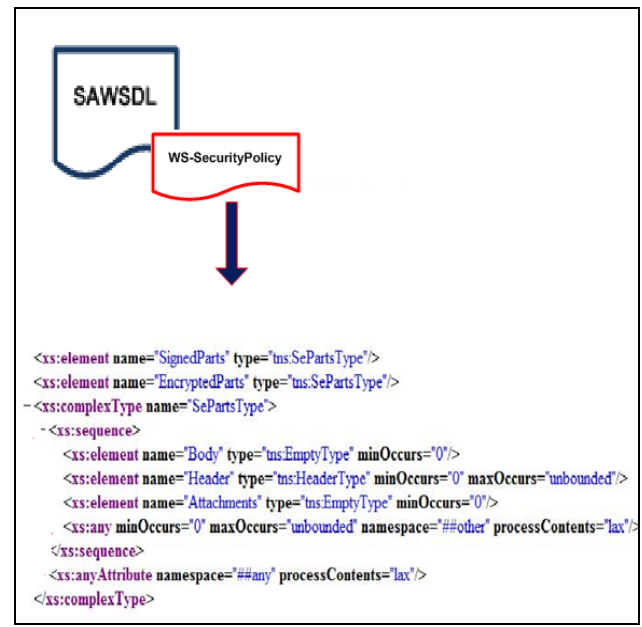

Figure 3. Enriched SAWSDL files by WS-SecurityPolicy tags

The proposed security solution has been mainly designed for providing authentication and encryption mechanisms for distributed systems. It is based on Public Key Infrastructure (PKI) certification. Certificate authority provides digital certificates and support managing, issuance and revocation of secured certificates using PKI. It is also called the trusted third part. Indeed, the e-health participants may use public and private keys to encrypt and decrypt the exchanged SOAP messages. Figure 4 illustrates the keys needed by both the sender and receiver to encrypt as well as decrypt a SOAP message. Both encrypting and decrypting operations need using receiver or sender public key before receiving or sending SOAP messages. Digital certificates insure message identity and confidentiality (for example, when receiving drugs order, Regional Depot of Drugs' administrator will be able to identify the hospital pharmacy's administrator who sent the order, and to ensure the integrity of order's data). In this context, it is not necessary to await the receipt of manually signed orders.

Being based on Web services, our architecture considers the use of *National e-Health Services Registry* which is a repository storing meta-data concerning all developed e-health services supported by the health community. Therefore, this registry is expected to fit with the specifications of the Universal Description Discovery and Integration (UDDI) repository version 3 [15]. To provide maximum security, runtime access to National e-Health Services Registry server must be available for medical stuff and patients through PKI certifications. This will ensure a high level of communication security.

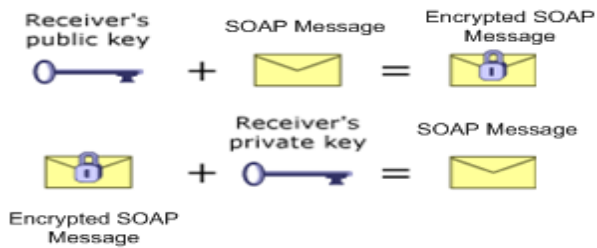

Figure 4: Encryption of SOAP Message

## V. CONCLUSION

In this paper, we presented main components of a distributed Moroccan HPIS which overcome the current system limitations and fit with a global Web services-based e-health architecture. Focusing on such a system is motivated by the real benefits of efficient e-health processes for Moroccan population, in particular, medications supply and consumption processes. The aim of this work is to overtake existing issues due to the heterogeneity and multiplicity of medical systems and data resources involved in these processes, and enable future collaboration with other e-health services. Our main contribution consists in elaborating Web services-based e-health architecture and describing its key components. The proposed architecture considers security techniques using PKI certificates, SOAP messages encryption and secured access to Web services registry to prevent possible malicious actions. It also ensures semantic interoperability through e-health ontology. Besides, the proposed National and Regional Drugs' order services provide a flexible solution to integrate all medications supply's sub-processes, when Prescription Data Integration Service is proposed to allow dynamic prescription data integration from various data sources.

### AUTHORS PROFILE

**Hajar Omrana** is currently pursuing PhD in Computer Science and works as IT consultant in a software services company. She received a degree in Computer Science in 2003 and DESA degree in Computing Networks, Telecommunications and Multimedia in 2006. She is also a member of IEEE Computer Society. Ten of her papers were published between 2010 and 2011. Her current research interests are semantic web services composition, quality of web services and MDA.

**Safae Nasiri** is currently pursuing PhD in Computer Science and works as IT consultant in Moroccan Ministry of Health. She received a degree in Computer Science in 2008. Her current research interests are Business Intelligence and e-Health systems.

**Fatima-Zahra Belouadha** is a Professor at Computer Science Department, EMI. She received a PhD in Computer Science in 1999. She received best paper award at SIIE'08 Conference and recognition award at IEEE International AICCSA'09 Conference. More than twenty of her papers were published between 2007 and 2011. Her current research interests are semantic composite web services, pervasive information systems/m-services, business intelligence and MDA.

**Ounsa Roudiès** is a Professor and Chief of Computer Science Department, EMI. She received doctorate degree in Computer Science in 1989 and PhD in Computer Science in 2001. She is co-editor of e-ti journal. More than thirty of her papers were published between 2008 and 2011. Her current research interests are SI, composition, web services, patterns and quality.